\documentclass[prb,aps,eqsecnum]{revtex4}
\usepackage{graphicx}

\newcommand{\ct}{{\cal T}}
\newcommand{\ep}{\epsilon}

\newcommand{\be}{\begin{equation}}
\newcommand{\ee}{\end{equation}}
\newcommand{\ba}{\begin{eqnarray}}
\newcommand{\ea}{\end{eqnarray}}

\newcommand{\la}{\label}

\newcommand{\dt}{\Delta t}

\begin{document}
\title{Any order imaginary time propagation method for solving the
Schr\"odinger equation}

\author{ Siu A. Chin$^\dagger$, S. Janecek$^\ddagger$ and
E. Krotscheck$^\ddagger$}
\affiliation{$^{\dagger}$Department of Physics, 
Texas A\&M University, College Station, TX 77843, USA}
\affiliation{
$^{\ddagger}$Institut f\"ur Theoretische Physik, Johannes Kepler
Universit\"at Linz, A-4040 Linz, Austria}

\begin{abstract}

  The eigenvalue-function pair of the 3D Schr\"odinger equation can be
  efficiently computed by use of high order, imaginary time
  propagators. Due to the diffusion character of the kinetic energy
  operator in imaginary time, algorithms developed so far are at most
  fourth-order. In this work, we show that for a grid based algorithm,
  imaginary time propagation of any even order can be devised on the
  basis of multi-product splitting. 
  The effectiveness of these
  algorithms, up to the 12$^{\rm th}$ order, is demonstrated by
  computing all 120 eigenstates of a 
  model C$_{60}$ molecule to very high precisions. 
  The algorithms are particularly useful when implemented on
  parallel computer architectures.

\end{abstract}
\maketitle

\section {Introduction}

With the advance of the density functional method of solving diverse
solid state physics and quantum chemistry problems\cite{torsti06}, it
is of growing importance to solve the Schr\"odinger equation on a
large 3-D mesh with greater than $N=10^6$ grid points. For such a
large mesh, conventional matrix methods are impractical, since even
the minimal matrix-vector multiplication would be prohibitively
slow. Among ${\cal O}(N)$ methods, we have previously shown that
fourth-order imaginary time propagation\cite{auer01}
provides an effective means of solving the Kohn-Sham
and related equations\cite{aich05}. The use of all {\it forward\/} time-step
fourth-order algorithms in solving the imaginary time Schr\"odinger
equation has since been adapted by many research
groups\cite{pil06,leh07,ren08}.

The lowest $n$ states of the one-body Schr\"odinger equation
\begin{equation}
H\psi_j({\bf r}) = E_j\psi_j({\bf r})
\end{equation}
with Hamiltonian
\begin{equation}
H= -{\hbar^2\over 2m}\nabla^2 + V({\bf r})\equiv T+V
\label{Ham}
\end{equation}
can be obtained in principle by applying the
evolution operator ($\epsilon=-\Delta t$)
\begin{equation}
{\cal T}(\epsilon)  \equiv e^{\epsilon(T+V)}
\label{evol}
\end{equation}
repeatedly on the $\ell$-th time step
approximation  $\bigl\{\psi_j^{(\ell\,)}({\bf r})\bigr\}$
to the set of states $\left\{\psi_j({\bf r}),\
1\le j \le n\right\}$,
\begin{equation}
\phi_j^{(\ell+1)} \equiv {\cal T}(\epsilon) \psi_j^{(\ell+1)}\label{eq:propstep}
\end{equation}
and orthogonalize the states
after every step,
\begin{equation}
\psi^{(\ell+1)}_j \equiv \sum_i c_{ji}\phi_i^{(\ell+1)},
\quad \bigl(\psi^{(\ell+1)}_j\bigr|\psi^{(\ell+1)}_i\bigr) = \delta_{ij}.
\label{eq:ortho}
\end{equation} The method is made
practical by approximating the exact evolution operator (\ref{evol})
by a general product form, 
\begin{equation}
{\cal T}(\epsilon)
= \prod_{i=1}^M e^{a_i\,\epsilon  T}e^{b_i\,\epsilon  V}.
\label{fact}
\end{equation}
The simplest second order decomposition, or the split operator
method\cite{feit2}, is \be {\cal T}_2(\epsilon)\equiv e^{{1\over
    2}\epsilon V} e^{\epsilon T} e^{{1\over 2}\epsilon V} = {\cal
  T}(\epsilon) + {O}(\epsilon^3).  \la{so} \ee When this operator acts
on a state $\psi_j({\bf r})$, the two operators $e^{{1\over 2}\epsilon
  V}$ correspond to point-by-point multiplications and $e^{\epsilon
  T}$ can be evaluated by one complete (forward and backward) Fast
Fourier Transform (FFT). Both are ${\cal O}(N)$ processes.  For ${\cal
  T}_2(\epsilon)$, $|\ep|$ has to be small to maintain good accuracy
and many iterations are therefore needed to project out the lowest $n$
states. To achieve faster convergence, one could in principle iterate
higher order algorithms at larger time steps. Unfortunately,
Sheng\cite{sheng89} and Suzuki\cite{suzuki91} have proved that, beyond
second order, no factorization of the form (\ref{fact}) can have all
positive coefficients $\{a_i,b_i\}$. This forward time step
requirement is essential for imaginary time propagation because if any
$a_i$ were negative, then the operator $e^{-a_i\dt T}$ would be
unbounded, resulting in unstable algorithms corresponding to
unphysical backward diffusion in time.  To derive forward, all
positive time step fourth-order algorithms, Suzuki\cite{Suzuki96} and
Chin \cite{ChinPLA97} have shown that a correction to the potential of
the form $\left[V,\left[T,V\right]\right]=(\hbar^2/ m)|\nabla V|^2$,
as first used by Takahashi-Imada\cite{ti84} and later suggested by
Suzuki\cite{su95}, must be included in the decomposition process. We
have shown\cite{auer01} previously that these forward fourth-order
algorithms can achieve similar accuracy at an order-of-magnitude
larger step sizes than the second-order splitting (\ref{so}).  More
recently Bandrauk, Dehghanian and Lu\cite{band06} have suggested that,
instead of including such a gradient term, one can use complex
coefficients $\{a_i,b_i\}$ having positive real parts. For real time
propagation, their complex time-step algorithms are not left-right
symmetric and therefore are not time-reversible. For imaginary time
propagation, their fourth-order algorithm $S^{\,\prime}_4$ requires
{\it five} complete complex-to-complex FFTs, whereas our forward
algorithm 4A only needs two
real-to-complex/complex-to-real FFTs\cite{auer01}.

\section {Multi-product expansion}

If the decomposition of ${\cal T}(\epsilon)$ is restricted to a single
product as in (\ref{fact}), then there is no practical means of
implementing a sixth or higher order forward algorithm\cite{chin051}.
However, if this restriction is relaxed to a sum of products, \be {\rm
  e}^{\ep( T+ V)}=\sum_k c_k \prod_{i} {\rm e}^{a_{k,i}\ep T}{\rm
  e}^{b_{k,i}\ep V}
\label{mprod} 
\ee then the requirement that $\{a_{k,i},b_{k,i}\}$ be positive means
that each product can only be second order. Since ${\cal T}_2(\ep)$ is
second order with positive coefficients, its powers ${\cal
  T}_2^k(\ep/k)$ can form a basis for such a multi-product
expansion. Recent work\cite{chin083} shows that such an expansion is
indeed possible and takes the form

\be
{\rm e}^{\ep( T+ V)}=\sum_{k=1}^n c_k {\cal T}_2^k\left(\frac\ep{k}\right)
+O(\ep^{2n+1})
\la{mcomp}
\ee
where the coefficients
$c_k$ are given in closed form for any $n$: 
\be
c_i=\prod_{j=1 (\ne i)}^n\frac{k_i^2}{k_i^2-k_j^2}
\la{coef}
\ee
with $\{k_1, k_2, \dots k_n\}=\{1,2, \dots n\}$.
Since the symmetric ${\cal T}_2(\ep)$ has only odd powers
in $\epsilon$,
\be
{\cal T}_2(\ep)=\exp[\ep (T+V)+\ep^3 E_3+\ep^5 E_5+\cdots]
\la{secerr}
\ee
where $E_i$ are higher order commutators of $T$ and $V$,
the expansion (\ref{mcomp}) is just a systematic extrapolation
which successively removes each odd order error.
Explicitly, for $n=2$ to 5, we have the following 
order 4 to order 10 multi-product expansion:
\be
{\cal T}_4(\ep)=-\frac13{\cal T}_2(\ep)
+\frac43{\cal T}_2^2\left(\frac\ep{2}\right)
\la{four}
\ee
\be
{\cal T}_6(\ep)=\frac1{24} {\cal T}_2(\ep)
-\frac{16}{15}{\cal T}_2^2\left(\frac\ep{2}\right)
+\frac{81}{40}{\cal T}_2^3\left(\frac\ep{3}\right)
\la{six}
\ee
\be
{\cal T}_8(\ep)=-\frac1{360} {\cal T}_2(\ep)
+\frac{16}{45}{\cal T}_2^2\left(\frac\ep{2}\right)
-\frac{729}{280}{\cal T}_2^3\left(\frac\ep{3}\right)
+\frac{1024}{315}{\cal T}_2^4\left(\frac\ep{4}\right)
\la{eight}
\ee
\be
{\cal T}_{10}(\ep)=\frac1{8640} {\cal T}_2(\ep)
-\frac{64}{945}{\cal T}_2^2\left(\frac\ep{2}\right)
+\frac{6561}{4480}{\cal T}_2^3\left(\frac\ep{3}\right)
-\frac{16384}{2835}{\cal T}_2^4\left(\frac\ep{4}\right)
+\frac{390625}{72576}{\cal T}_2^5\left(\frac\ep{5}\right).
\la{ten}
\ee
Since each $\ct_2$ requires one complete FFT, the above series of
$2n$-order algorithms only requires $n(n+1)/2$ complete FFTs. Thus
algorithms of order 4, 6, 8 and 10 only require 3, 6, 10 and 15
complete FFTs. The low order extrapolation (\ref{four}) has been 
used previously\cite{kevin95}. Here,
we have a systematic expansion to any even order. Note that 
Romberg-type extrapolation\cite{kevin95} such as
\be
{\cal T}_6(\ep)=-\frac1{15}{\cal T}_4(\ep)
+\frac{16}{15}{\cal T}_4^2\left(\frac\ep{2}\right)
\la{sixp},
\ee
which triples the number of FFTs in going from order $2n$ to 
$2n+2$, is not competitive 
with Eq. (\ref{mcomp})'s linear increase of only $n+1$ additional FFTs.

Since some coefficients $c_k$ are negative, this requires that the
corresponding product, when acting on state $\psi_j$, be subtracted.
This is doable for a grid based discretization of the wave function,
which is just a point-by-point subtraction.

\section {Quantum well model of C60}

To demonstrate the workings of this new family of algorithms, we apply
them to a model potential with nontrivial geometry, that of a 3D
C$_{60}$ molecule. The effective attraction of the carbon ions
is modeled by a potential of the form 
\be V({\bf r}) = -\sum_i \frac{V_0}{\cosh(|{\bf r}-{\bf R}_i|/d)} 
\ee 
where ${\bf R}_i$ are the locations of the carbon atoms in the C$_{60}$ cage.
The
strength $V_0$ was chosen $1$ in units of $\hbar^2/2m$ and the width
of the troughs $d$ = 0.05 a.u.  This potential accommodates 120 bound
states as needed for a C$_{60}$ calculation. We have also applied
the method in 2D to a square grid of 9$\times$9 quantum dots
described by the same potential.
The convergence behavior of the algorithms in both cases are practically
identical and need not be discussed separately.

While the above potential is not a realistic description of a C$_{60}$
molecule, it serves to highlight an important and realistic aspect of
any such computation: In the 3D case, the lowest 120 eigenvalues
consist of two groups of 60 almost degenerate eigenvalues, one
centered around an energy of -5.245~$\hbar^2/2m$ with an average level
spacing of 0.0034~$\hbar^2/2m$, the other one around
-2.992~$\hbar^2/2m$ with an average level spacing of
0.0077~$\hbar^2/2m$.  Degenerate eigenvalues pose a notorious problem
for eigenvalue solvers that contain an orthogonalization step. In this
implementation of the algorithm, we use the subspace orthogonalization
method described in Ref.~\onlinecite{aich05},
which is an application of the Petrov-Galerkin method\cite{torsti06}.
 We again find that the
method works well in the present case; the convergence rates for the
highest and the lowest states are the same.

\begin{figure}
	\centerline{\includegraphics[width=0.75\linewidth]{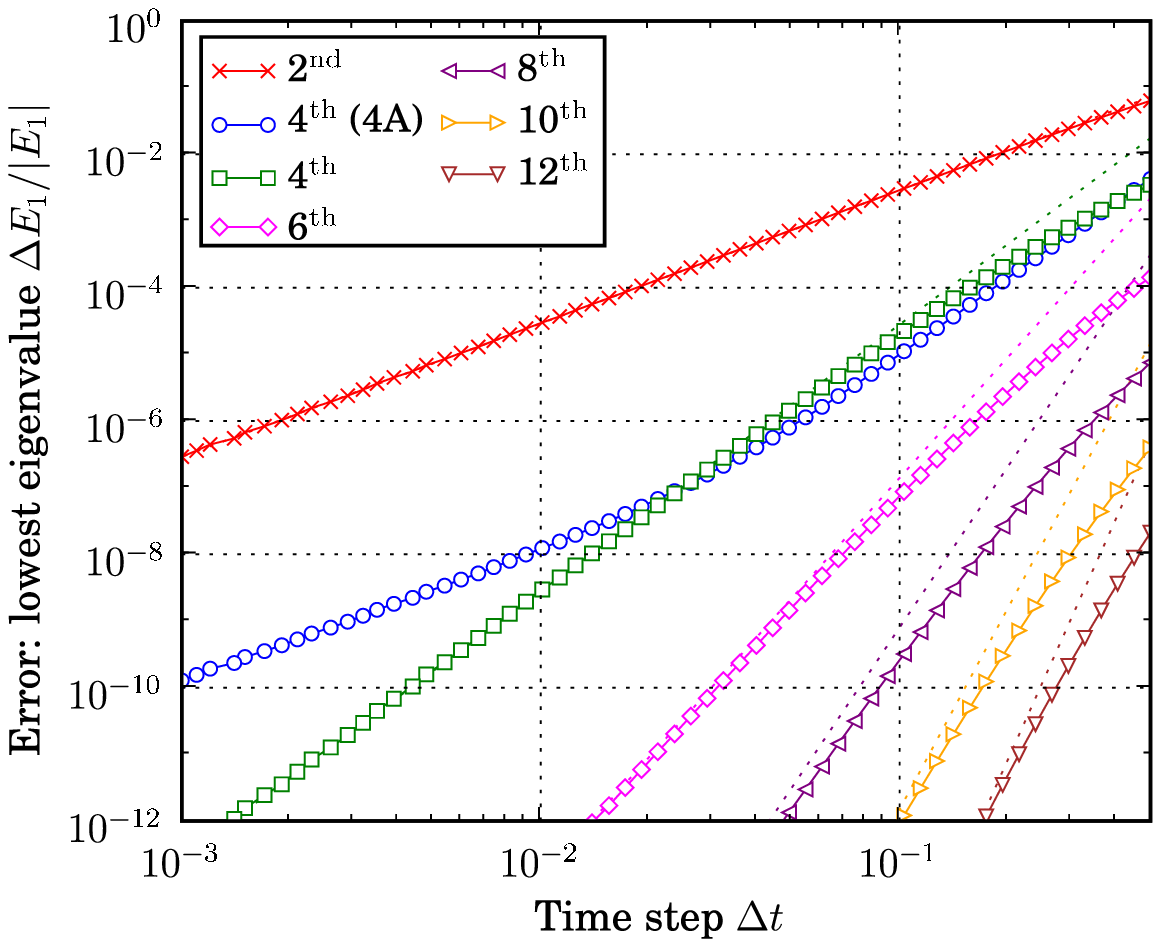}}
\caption{(color online)
The convergence of the 2$^{\rm nd}$ to 12$^{\rm th}$ order algorithms
for the lowest eigenstate of a model ``C$_{60}$'' molecule
are as shown by
markers defined in the inset. The dashed lines, as a guide to
the eye, are the power laws $\dt^n$ for $n = 2, 4, \ldots, 12$. Also
shown is the convergence curve for algorithm 4A of Ref. \onlinecite{auer01}.
Its characteristic deviation from the $\dt^4$ power law behavior at
small $\dt$
is due to discretization errors in evaluating the
double commutator $\left[V,\left[T,V\right]\right]$. 
\label{fig:c60one}}
\end{figure}

\begin{figure}
	\centerline{\includegraphics[width=0.75\linewidth]{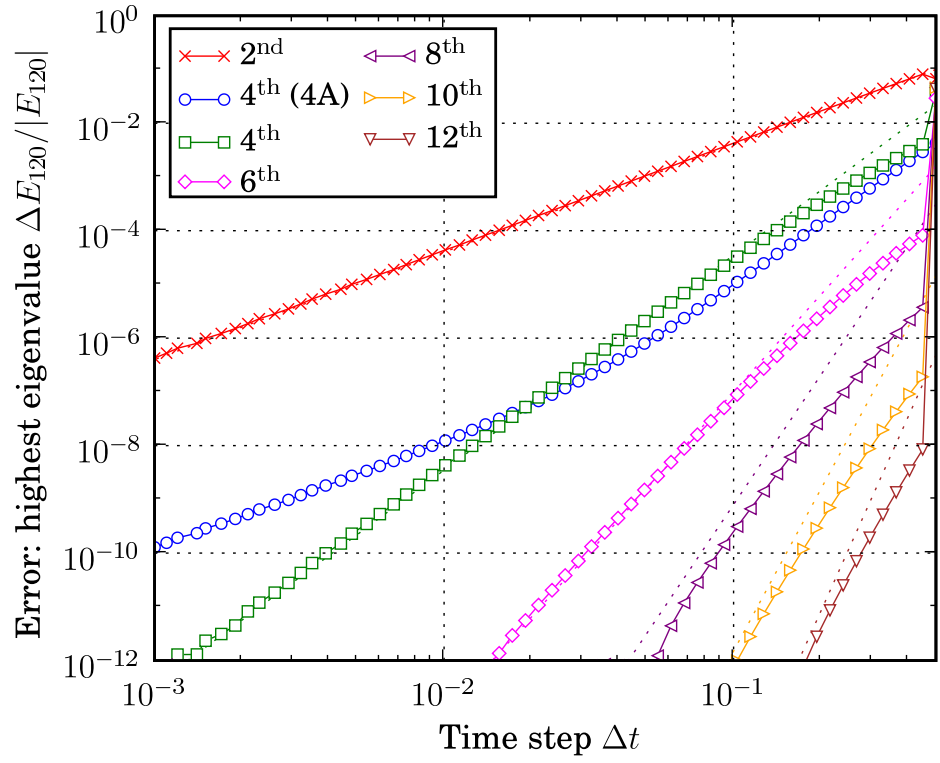}}
\caption{
Same as Fig. \ref{fig:c60one} for the 120$^{\rm th}$ 
eigenstate of the model
C$_{60}$ molecule. 
\label{fig:c60120}}
\end{figure}

Figs.~\ref{fig:c60one} and \ref{fig:c60120} show the convergence of
various algorithms for both the lowest and the highest state of the
model C$_{60}$ molecule. To generate the figures, we started with a
time step of $\dt=0.5$ and a set of plane wave initial states in an
appropriate box. We then reduced the time step by a factor of 0.9 each
time after convergence has been reached, so that the power law
behavior can be seen cleanly. In a realistic calculation, it suffices
to reduce the time step by a factor 0.5. In other words,
the 12$^{\rm th}$ order algorithm can reach the 10$^{-10}$ error level
in just two iterations. The reduction factor of 0.5 used here is empirical.
Recently, Lehtovaara, Toivanen and Eloranta\cite{leh07} have suggested that
the time step size can be optimally adjusted with added efforts.

While the convergence rate of the eigenvalues
verified the order of the algorithms, 
in practice, it is also useful to monitor the
variance of all states with respect to the evolution operator,
\be
R_j^E = \sum_k\left|{\cal T}_n(\ep) \psi_j({\bf r}_k)
- e^{\ep E_i} \psi_j({\bf
  r}_k) \right|^2
\ee
Only states with $R_j^E>\gamma$, where $\gamma$ is a
prescribed error bound, need to be propagated and orthogonalized. As
soon as all states have converged at a certain time step $\ep$, their
variances with respect to the Hamiltonian,
\be R_j^H =\sum_k\left| H
\psi_j({\bf r}_k) - E_i \psi_j({\bf r}_k)\right|^2,
\ee
are calculated. If
$R_j^H< \gamma$ for all states $j$, the iterations are terminated,
otherwise the time step is reduced and the whole process is repeated,
taking the result of the previous iteration as initial values.

\section{Parallelization}

The advantage of high-order propagation methods is particularly
compelling on parallel computer architectures: The propagation step
(\ref{eq:propstep}) can be parallelized efficiently without having to
abandon the advantage of using FFTs by simply distributing the states
$\psi_j$ across different processors. In such an arrangement, however,
the parallelization of the orthogonalization step is notoriously
difficult.  Let $T_\mathrm{pro}$ be the propagation time, {\it i.e.\/},
the time it takes to carry out step (\ref{eq:propstep}) for all
states, and $T_\mathrm{ort}$ the orthogonalization time. Then, the
time $T_\mathrm{tot}$ for one iteration step on an ideal machine with
$N$ processors is
\begin{equation}
T_\mathrm{tot}(N) = T_\mathrm{pro}/N + T_\mathrm{ort},
\end{equation}
and the speed up ratio for the ``propagation only'' and the
total time
step including orthogonalization for the $j^{\rm th}$ order algorithm is
\begin{equation}
S_{\mathrm{pro/tot}}^{(j)}(N) = 
\frac{T_\mathrm{pro/tot}(1)}{T_\mathrm{pro/tot}(N)}\,,
\end{equation}
assuming the number of states is larger than the number of processors
allocated for the task. 
The actual speed-up ratio will be less than this ideal
since we have neglected communication overhead and other hardware/system
specific issues.

\begin{figure}
	\centerline{\includegraphics[width=0.75\linewidth]{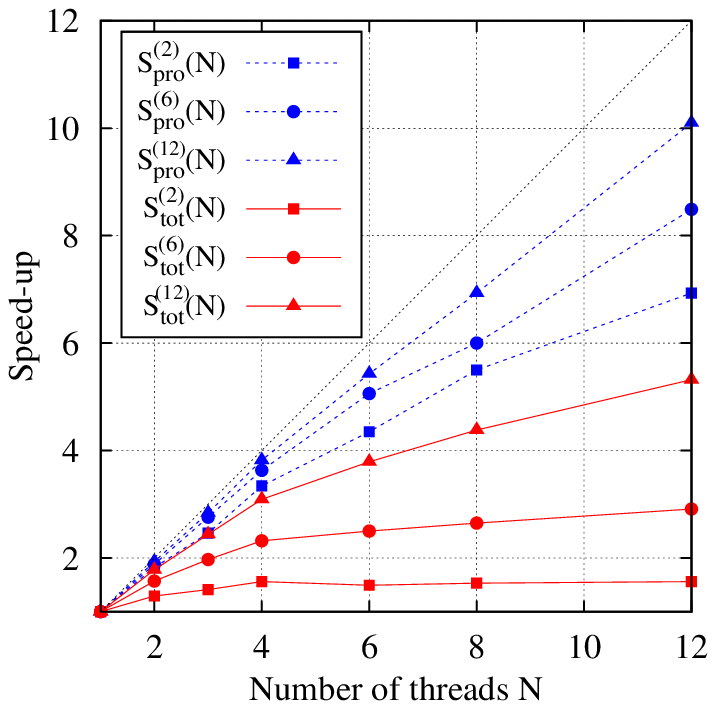}}
\caption{(color online)
The total speed-up time factor $S_{\rm tot}$ (solid lines) 
and the propagation only (without orthogonalization) time
speed-up factor $S_{\rm pro}$, as a function of
parallel threads,
for the 2$^{\rm nd}$, 6$^{\rm th}$, and 12$^{\rm th}$ order
algorithm (filled squares, circles, and triangles, respectively).
Also shown is the ``ideal'' speed-up factor (dotted line).
\label{fig:speedup}
}
\end{figure}

Fig. \ref{fig:speedup} shows the speed up ratio for the C$_{60}$ model
calculation
in the case of the 2$^{\rm nd}$, 6$^{\rm th}$, and 12$^{\rm th}$
order algorithms on a 256 Itanium\cite{Intel} processor
Altix\cite{SGI} machine for up to twelve threads. We show 
the two speed-up ratios 
$S_{\mathrm{pro}}^{(j)}(N)$ and
$S_{\mathrm{tot}}^{(j)}(N)$. No particular effort was made to
parallelize the orthogonalization step. Evidently, the speed-up of the
propagation step alone is a reasonably linear function of the number
of threads. The performance improves significantly with the order of
the algorithm because the increase computational effort for propagation
can be distributed while the cost of communication remained the same.

The 12$^{\rm th}$ order algorithm can reach
about 80 percent of the optimal performance.  The actual speed-up is
limited by the orthogonalization step; while the 12$^{\rm th}$ order
algorithm can still attain a more than five-fold speed-up, it is
hardly worth parallelizing the second order algorithm.
The specific speed-up factor for higher order algorithms also depends 
on the number $n$ of needed eigenstates . In general, the time for 
propagation is essentially 
proportional to $n$, 
whereas the time for
orthogonalization goes as $n^2$.

Thus, high order algorithms provide two advantages: 
they have 
faster convergence at larger time steps
and are more adaptive to parallel computing environments. In the single 
processor mode, we have determined that
the 6$^{\rm th}$ order algorithm 
performed the best.

\section{Conclusions}

The impressive convergence of our high--order algorithms has a
computing cost: One propagation step of the $2n$-th order algorithm is
equivalent to $n(n+1)/2$ propagation steps of the second order
algorithm. This cost is compensated by two effects: The first is shown
in the figures: The much faster convergence as a function of
time step implies that fewer iterations are needed to complete the
calculation. The second advantage is less obvious; since
orthogonalization is carried out {\it after\/} the propagation step
(\ref{mcomp}), the relatively costly number of orthogonalization steps is
dramatically 
reduced.

The most likely use of the high-order algorithms will be in real-space
implementations of density-functional theory. For realistic systems,
one must include non-local pseudo-potentials.  We have recently
\cite{nonloc} implemented algorithm 4A using pseudo-potentials of the
Kleinman-Bylander form \cite{Klei82}. Calculating the double commutator
$\left[V,\left[T,V\right]\right]$ for such non-local potentials is
possible, but the computational cost is twice that of propagating
just the potential. 
Moreover, if the electron density in the vicinity of 
the ion cores deviates from spherical symmetry, then some
approximate treatments may degrade the order of the algorithm. 
Thus, our new algorithms,
without needing the double commutator, should be even more 
effective for realistic density-functional calculations.

\begin{acknowledgments}
This work was supported, in part, by the Austrian Science Fund FWF
(to EK) under project P18134.
\end{acknowledgments}
\newpage
\centerline{REFERENCES}

\newpage




\begin{thebibliography}{}

\bibitem{torsti06} T. Torsti {\rm et al.\/}, Physica Status Solidi {\bf B 243}
  (2006) 1016.

\bibitem{auer01} J. Auer, E. Krotscheck and S. A. Chin,
        J. Chem. Phys. {\bf 115} (2001) 6841. 

 \bibitem{aich05} M. Aichinger and E. Krotscheck, Computational Material
   Science {\bf 34}, (2005) 188. 

 \bibitem{pil06} L. Brualla, K. Sakkos, J. Boronat and J. Casulleras,
   J. Chem. Phys. {\bf 121} (2004) 636. 
		
 \bibitem{leh07} L. Lehtovaara, J. Toivanen, and J. Eloranta, J. Comp. Phys. 
            {\bf 221}, (2007) 148. 

 \bibitem{ren08} G. B. Ren and J. M. Rorison, Phys. Rev. {\bf B 77}
   (2008) 245318. 

\bibitem{feit2}
D. Feit, J.~A. {Fleck, Jr.}, and A. Steiger, J. Chem. Phys. {\bf 78} 
   (1982), 301.

\bibitem{sheng89} Q. Sheng, 
          IMA J. Numer. Analysis, {\bf 9} (1989) 199 .

\bibitem{suzuki91} M. Suzuki, 
		  J. Math. Phys. {\bf 32} (1991) 400.

\bibitem{Suzuki96}
M. Suzuki,  in {\em Computer Simulation Studies in Condensed Matter Physics},
  edited by D.~P. Landau, K.~K. Mon, and H.-B. Sh{\"u}ttler (Springer, Berlin,
  1996), Vol.~VIII.

\bibitem{ChinPLA97}
S.~A. Chin, Phys. Lett. A {\bf 226} (1997) 344.

\bibitem{ti84}M. Takahashi and M. Imada,
			  J. Phys. Soc. Jpn., {\bf 53} (1984) 3765.
\bibitem{su95}
M. Suzuki, Phys. Lett. A {\bf 201} (1995) 425.

\bibitem{band06} A. D. Bandrauk, E. Dehghanian and H. Lu,
Chem. Phys. Lett {\bf 419} (2006) 346.

\bibitem{chin051} S. A. Chin,
Phys. Rev. {\bf E 71} (2005) 016703 . 

\bibitem{chin083} S. A. Chin, ``Multi-product Splitting and
  Runge-Kutta-Nystr\"om Integrators'', arXiv:0809.0914.

\bibitem{kevin95} K. E. Schmidt and M. A. Lee, Phys. Rev. {\bf E 51}
  (1995) 5495. 

\bibitem{Intel}
Itanium is a registered trademark of Intel Corp.

\bibitem{SGI}
Altix is a registered trademark of Silicon Graphics inc.	 

\bibitem{Lapack}
{\tt www.netlib.org/lapack}.

\bibitem{nonloc} E.~R.~Hern\'{a}ndez, S.~Janecek,
     M.~S.~Kaczmarski, and E.~Krotscheck",
     Phys. Rev. {\bf B 75} (2007)  075108.

\bibitem{Klei82}
L. Kleinman and  D.~M. Bylander,
   Phys. Rev. Lett. {\bf 48}  (1982) 1425.

\end{thebibliography}
\end{document}